\documentclass[aps,preprint,showkeys,superscriptaddress,floatfix]{revtex4}

\usepackage{graphicx}		% Include figure files
\usepackage{bm}				% bold math
\usepackage{amsmath}
\usepackage{amssymb}
\usepackage{amsfonts}
\usepackage{amssymb}
\usepackage{color}
\usepackage{enumitem}
\usepackage{hhline}

\begin{document}

\title{Confined magnetoelastic waves in thin waveguides}

\author{Frederic Vanderveken}
\email[E-mail: ]{frederic.vanderveken@imec.be}
\affiliation{Imec, 3001 Leuven, Belgium}
\affiliation{KU Leuven, Departement Materiaalkunde, SIEM, 3001 Leuven, Belgium}

\author{Jeroen Mulkers}
\author{Jonathan Leliaert}
\author{Bartel Van Waeyenberge}
\affiliation{Universiteit Gent, Departement Vastestofwetenschappen, DyNaMat, 9000 Gent, Belgium}

\author{Bart Sor\'ee}
\affiliation{Imec, 3001 Leuven, Belgium}
\affiliation{KU Leuven, Departement Elektrotechniek, TELEMIC, 3001 Leuven, Belgium}
\affiliation{Universiteit Antwerpen, Departement Fysica, 2000 Antwerpen, Belgium}
\author{Odysseas Zografos}
\affiliation{Imec, 3001 Leuven, Belgium}
\author{Florin Ciubotaru}
\affiliation{Imec, 3001 Leuven, Belgium}
\author{Christoph Adelmann}
\email[E-mail: ]{christoph.adelmann@imec.be}
\affiliation{Imec, 3001 Leuven, Belgium}

\begin{abstract}
The characteristics of confined magnetoelastic waves in nanoscale ferromagnetic magnetostrictive waveguides have been investigated by a combination of analytical and numerical calculations. The presence of both magnetostriction and inverse magnetostriction leads to the coupling between confined spin waves and elastic Lamb waves. Numerical simulations of the coupled system have been used to extract the dispersion relations of the magnetoelastic waves as well as their mode profiles. 
\end{abstract}


\maketitle

\section{Introduction}

In recent years, the coupling between elastic and magnetic degrees of freedom in magnetostrictive materials has gained renewed interest due to emerging nanoscale spintronic applications, such as magnetic memory cells \cite{kent_new_2015,Kawahara2012,Klimov17,Chen19}, logic devices \cite{Verba19,Sadovnikov19,Chumak2015,Dieny2020}, sensors \cite{Nan85,Dong06}, or compact microwave antennas \cite{Yao20,Schneider19}. In magnetostrictive ferromagnets, the magnetoelastic coupling leads to an elastic response to a magnetic excitation and \emph{vice versa}. At microwave frequencies, this mutual interaction manifests itself in a coupling between spin waves---the fundamental magnetic excitations in this frequency range---and (hypersonic) elastic waves, forming magnetoelastic waves. While this behavior has been studied for plane waves in bulk materials decades ago \cite{Kittel58,Schlomann60,Schlomann64,Rezende69,Kobayashi73,Fedders74,Gurevich96,Tucker1972}, magnetoelastic waves in nm-thin films have been studied only much more recently. Most of these studies have focused on the interaction between surface acoustic waves propagating at the interface between a (piezoelectric) substrate and a thin magnetostrictive film with macroscopic dimensions \cite{Verba19,Duquesne19,Chang18,Verba18,NOZAKI18,Li17,Bhuktare17,Gowtham16,Labanowski16,Dreher12,Puebla20,Weiler11,Sasaki17,Castilla20,Zhou14,Ganguly76}. In addition, also numerical and theoretical studies on magnetoelastic plane waves in bulk media have been reported \cite{Chen17,Graczyk17,Graczyk172,Ruckriegel14,Streib19}.

By contrast, many recent spin-wave-based information processing applications employ nanoscale ferromagnetic waveguides for information transfer and computation \cite{Khitun11,Mahmoud20,Fischer17,Talmelli19}. The small dimensions of such waveguides, which are required to enable high device densities, lead to wave confinement effects when the wavelengths become comparable to the waveguide width. In the presence of (inverse) magnetostriction, this leads to the coupling between confined spin waves and confined elastic waves, forming \emph{confined} magnetoelastic waves. To date, studies have addressed the effect of confinement on spin waves \cite{Vanderveken20,Wang19,Demidov15,Guslienko02,Guslienko05,Guslienko11} as well as elastic waves \cite{Jean14,Mante13,Tadesse15,Jean15,Belliard13,Ristow13,Xie19,Safavi-Naeini19,Mante18}. Yet, besides recent investigations of the magnetoelastic coupling in nanoscale resonators \cite{Polzikova16,Ulrichs17,Polzikova18,Berk19,POLZIKOVA19}, a detailed study of \emph{propagating} magnetoelastic waves in nanoscale waveguides is still lacking. It is clear that a detailed understanding of confined propagating magnetoelastic waves is crucial for emerging magnonic device applications, especially where spin waves are excited by magnetoelectric means and used for information transfer and processing \cite{Khitun11,balinskiy_magnetoelectric_2018,Bozhko20,Mahmoud20}.

In this work, we report on a combined analytical and numerical description of the characteristics of confined magnetoelastic waves in thin and narrow waveguides. The numerical calculations employ a new mumax3 extension to solve the magnetoelastodynamics and allow for the assessment of confined magnetoelastic wave dynamics. The analytical model complements the numerical results and is utilized to gain insight in the coupling between the different confined elastic and magnetic modes. Thus, this work provides a key step towards the comprehensive understanding of confined magnetoelastic waves in nanoscale ferromagnetic waveguides.

\section{Theoretical description of magnetoelastic waves}

\subsection{Spin waves, Lamb waves, and magnetoelasticity\label{sec:theory}}

Our description of the magnetoelastic dynamics starts from the total energy density, given by
\begin{equation}
\label{eq:E_tot}
	\mathcal{E}_\mathrm{tot} = \mathcal{E}_\mathrm{Z} + \mathcal{E}_\mathrm{d} + \mathcal{E}_\mathrm{ex} + \mathcal{E}_\mathrm{mel} + \mathcal{E}_\mathrm{el} + \mathcal{E}_\mathrm{kin} .
\end{equation}
Here, $\mathcal{E}_\mathrm{Z}$ represents the Zeeman energy density, $\mathcal{E}_\mathrm{d}$ the demagnetization energy density, $\mathcal{E}_\mathrm{ex}$ the exchange energy density,  $\mathcal{E}_\mathrm{mel}$ the magnetoelastic energy density, $\mathcal{E}_\mathrm{el}$ the elastic energy density, and $\mathcal{E}_\mathrm{kin}$ the kinetic energy density. The magnetic energy densities can be expressed by \cite{Kittel58,Schlomann60,Schlomann64}
\begin{eqnarray}
\label{eq:E_Zeeman}
\mathcal{E}_\mathrm{Z} &=& -\mu_0 M_s \left(\bm{m}\cdot\bm{H}_\mathrm{ext}\right),  \\
\label{eq:E_dip}
\mathcal{E}_\mathrm{d} &=& - \frac{\mu_0 M_s}{2} \left(\bm{m}\cdot\bm{H}_\mathrm{d}\right), \\
\label{eq:E_ex}
\mathcal{E}_\mathrm{ex} &=& A_\mathrm{ex} \left[ \left(\nabla m_x\right)^2+ \left(\nabla m_y\right)^2 +  \left(\nabla m_z\right)^2\right], 
\end{eqnarray}
\noindent with $\bm{m}=\bm{M}/M_\mathrm{s}$ the magnetization $\bm{M}$ normalized to the the saturation magnetization $M_\mathrm{s}$, $\mu_0$ the vacuum permeability, $\bm{H}_\mathrm{ext}$ the external magnetic field strength, $\bm{H}_\mathrm{d}$ the demagnetization field strength, and $A_{\mathrm{ex}}$ the exchange stiffness constant.

The magnetoelastic energy density for a material with cubic (or higher) crystal symmetry is given by \cite{Gurevich96,Tucker1972}
\begin{equation}
\label{eq:E_mel}
\mathcal{E}_\mathrm{mel} = B_1 \sum_i m_\mathrm{i}^2 \varepsilon_{ii} + B_2 \sum_{i\neq j} m_{i}m_{j} \varepsilon_{ij}\,.
\end{equation}
\noindent Here, $\bar{\varepsilon}$ is the strain tensor with components $\varepsilon_{ij}$ and $B_{1,2}$ are the magnetoelastic coupling constants.

In the linear elastic regime, Hooke's law is valid and the elastic energy density is given by \cite{Barber2004,Graff1975}
\begin{equation}
\label{eq:E_el}
	\mathcal{E}_\mathrm{el} =  \frac{1}{2} \bar{\sigma}:\bar{\varepsilon} = \frac{1}{2} \sum_{i,j,k,l} C_{ijkl}  \varepsilon_{kl} \varepsilon_{ij},
\end{equation}
\noindent with $\bar{\sigma}$ the mechanical stress tensor and $C_{ijkl}$ the stiffness constants. Finally, the kinetic energy density can be expressed as \cite{Achenbach1973,Nayfeh1995}
\begin{equation}
\label{eq:E_kin}
\mathcal{E}_\mathrm{kin} = \frac{\rho||\dot{\bm{u}}||^2}{2 },
\end{equation}
\noindent with $\rho$ the mass density and $\bm{u}$ the mechanical displacement. 

The minimization of the total energy $E_{\mathrm{tot}}$, \emph{i.e.} the minimization of the volume integral of the energy density $\mathcal{E}_\mathrm{tot}$, then allows to find the equilibrium state. Beyond equilibrium, the magnetization dynamics and magnetic excitations in the system are described by the Landau-Lifshitz-Gilbert (LLG) equation \cite{LANDAU92,Gilbert04}
\begin{equation}
\label{eq:llg}
	\dot{\bm{m}} = -\gamma_0 \bm{m} \times \bm{H}_\mathrm{eff} + \alpha \bm{m}\times \dot{\bm{m}},
\end{equation}
\noindent with $\gamma_0=\mu_0 \gamma$, $\gamma$ the absolute value of the gyromagnetic ratio, $\alpha$ the phenomenological Gilbert damping constant, and $\bm{H}_\mathrm{eff}$ the effective magnetic field strength, which is given by 
\begin{equation}
\label{eq:h_eff}
	\bm{H}_\mathrm{eff} = -\frac{1}{\mu_0} \frac{\delta E_\mathrm{tot}}{\delta \bm{M}} \hspace{10pt} \text{with} \hspace{10pt} E_\mathrm{tot} = \int\limits_{V}\mathcal{E}_\mathrm{tot}dV\,.
\end{equation}
\noindent The boundary conditions for the magnetization are 
\begin{equation}
\label{eq:M_boundary}
	\nabla\bm{_n} \cdot \bm{m} = 0,
\end{equation}
\noindent with $\bm{n}$ the normal to the surface.

The elastodynamic equation of motion is given by \cite{Graff1975,Achenbach1973}
\begin{equation}
\label{eq:elastodynamic}
\rho \ddot{\bm{u}}  + \eta \dot{\bm{u}}= \bm{f}_\mathrm{tot},
\end{equation}
\noindent with $\eta$ a phenomenological damping parameter and $\bm{f}_\mathrm{tot}$ the total body force acting on the material. This body force is given by \cite{Graff1975,Achenbach1973}
\begin{equation}
\label{eq:bf}
	\bm{f}_\mathrm{tot} = \nabla \cdot \frac{dE_\mathrm{tot}}{d \varepsilon_{ij}} \hspace{10pt} \text{or} \hspace{10pt} f_{\mathrm{tot},i} = \frac{\partial}{\partial x_{j}} \frac{\delta E_\mathrm{tot}}{\delta \varepsilon_{ij}}
\end{equation}
\noindent and has both elastic and magnetoelastic contributions. For small values, the displacement is related to the strain by
\begin{equation}
\label{eq:strain_displacement}
\bar{\varepsilon} = \frac{1}{2} \left( \nabla \mathbf{u} + \left(\nabla \mathbf{u} \right) ^T \right),
\end{equation}
\noindent which can be used to express the total body force as a function of the displacement $\bm{u}$. The mechanical boundary conditions at the surface are 
\begin{equation}
\label{eq:U_boundary}
	f_\mathrm{s} = \bar{\sigma} \cdot \bm{n},
\end{equation}
\noindent with $f_\mathrm{s}$ the traction force per unit surface.

\subsection{Linear magnetoelastic waves in thin waveguides}
\label{sec:theory2}

In this section, we discuss an analytical approach to the above equations of motion following previous work on magnetoelastic waves in thin films \cite{Kamenetskii21}. Although the system is not generally solvable for confined waves in a narrow waveguide, the results provide useful insight in the underlying physics and the dependence of the coupling on the symmetry of the waves. They are therefore complementary to the numerical results presented in the next section below.

In the analytical description, the waveguide is considered to be infinitely long in the propagation direction $\bm{\hat{x}}$ with free boundaries in the other two directions. The magnetic Neumann boundary conditions in Eq.~\eqref{eq:M_boundary} are strictly satisfied if the surrounding materials are nonmagnetic. By contrast, the mechanical free boundary conditions $\bar{\sigma} \cdot \bm{n}=0$ are satisfied if the waveguide is surrounded by vacuum. In practice, a good approximation is already obtained when the waveguide is surrounded by materials with much lower acoustic impedances. Hence, the results obtained in this work are \emph{e.g.} relevant for supported waveguides if the underlying substrate has a much lower acoustic impedance than the waveguide, which causes total reflection of elastic waves at the interface. 

The waveguide thickness $d$ is considered to be much smaller than the wavelength $\lambda$ of the magnetoelastic wave, \emph{i.e.}\ $kd\ll 1$, with $k = {2\pi}/{\lambda}$ the wavenumber. As a result, the dynamic magnetization and the displacement can be assumed to be uniform over the thickness and all partial derivatives with respect to the direction normal to the waveguide, $\bm{\hat{z}}$, vanish (\emph{i.e.} $\partial/\partial z=0$). For scaled waveguides, their width is however of the same order as the magnetoelastic wavelength. Therefore, mode formation occurs due to confinement in the lateral direction along $\bm{\hat{y}}$. 

\begin{figure}[t]
	\includegraphics[width=8cm]{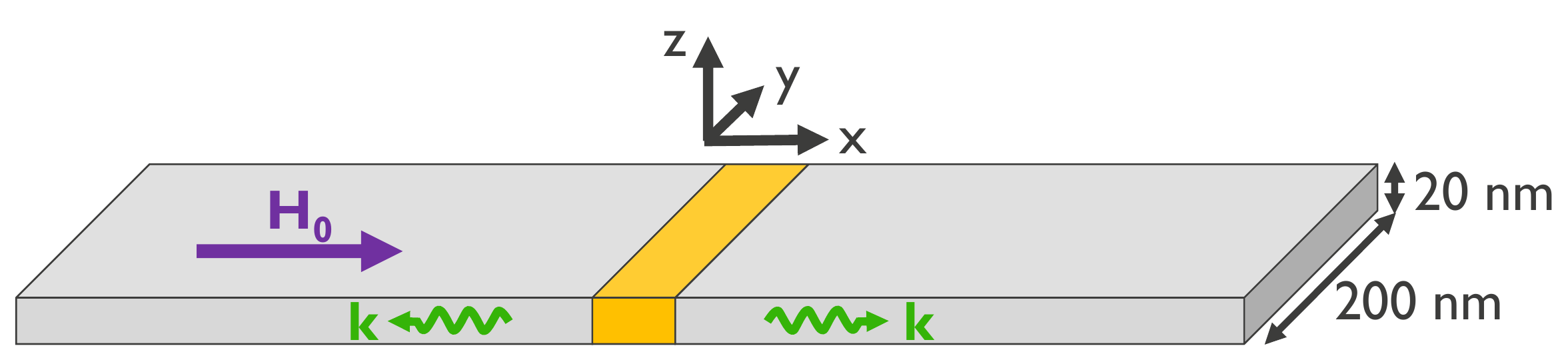}
	\caption{Schematic of the studied studied CoFeB waveguide (10 $\mu$m long, 200 nm wide, 20 nm thick). The yellow area designates the excitation region in the waveguide center. A static external magnetic field with an amplitude of $\mu_0H_0 = 5$ mT is applied along the waveguide in the $x$-direction.}
	\label{fig:geometry}
\end{figure}

A schematic of the waveguide geometry is shown in Fig.~\ref{fig:geometry} with the propagation direction and static external field along $\bm{\hat{x}}$. The state variable describing a propagating magnetoelastic wave can be written as
\begin{equation}
\label{eq:ansatz}
	\bm{w}_n(x,y,t) = \tilde{\bm{w}}_n(y) e^{i(k_xx+\omega t)} ,
\end{equation}
\noindent with $\tilde{\bm{w}}_n = [\tilde{u}_{x,n}, \tilde{u}_{y,n},\tilde{u}_{z,n},\tilde{m}_{y,n},\tilde{m}_{z,n}]^T$ and $n$ the mode number. Note that weak dynamic displacement and magnetization components are assumed. Hence, higher order terms are neglected and $m_x = 1$.

The amplitude and the profile $\tilde{\bm{w}}_n$ as well as the dispersion relation of magnetoelastic waves in the waveguide can be obtained by solving the coupled differential equations of motion \eqref{eq:llg} and \eqref{eq:elastodynamic}. A major complication is however the analytical self-consistent calculation of the demagnetization field. The problem can be considerably simplified by assuming that the confined spin-wave modes are not altered by the magnetoelastic interaction. As shown below, the exact numerical solutions of the coupled system indicate that this approximation is well justified. Then, the magnetization components in $\tilde{\bm{w}}_n$ can be written as \cite{Wang19,Demidov15}
\begin{equation}
\label{eq:mode_profile}
\tilde{m}_{i,n}(y) = A_{i,n} \begin{cases}
\cos(\kappa_n y)       & \quad \text{if } n \text{ is odd}\\
\sin(\kappa_n y)   	& \quad \text{if } n \text{ is even}
\end{cases}\,,
\end{equation}
\noindent with complex amplitudes $A_{i,n}$, $i \in \{ y,z\}$, and $\kappa_n$ the wavenumber along $\bm{\hat{y}}$. In a waveguide, $\kappa_n$ has discrete values of $\kappa_n = n\pi/w_\mathrm{eff}$, with $w_\mathrm{eff}$ the effective waveguide width \cite{Demidov15,Guslienko02,Guslienko05,Guslienko11} and $n$ the mode number. Note that no corresponding assumptions for the displacement components in $\tilde{\bm{w}}_n$ need to be made.

Using this approximation, neglecting damping, and further assuming small dynamic displacement and magnetization components, the coupled differential equations \eqref{eq:llg} and \eqref{eq:elastodynamic} can be linearized. Detailed calculations can be found in Appendix \ref{app:A}, which lead to a set of homogeneous partial differential equations
\begin{widetext}
\begin{equation}
\label{eq:lin_llg}
\begin{bmatrix}
v_\parallel^2k_x^2 - v_\perp^2 \partial_y^2 - \omega^2  &  -iv_\asymp^2k_x\partial_y  & 0 & \frac{i\kappa_nB_2}{\rho M_\mathrm{s}} & 0 \\
-iv_\asymp^2k_x\partial_y  & v_\perp^2 k_x^2 -v_\parallel^2 \partial_y^2 - \omega^2 & 0 & \frac{ik_xB_2}{\rho M_\mathrm{s}}  & 0 \\
0 & 0 & v_\perp^2 \left( k_x^2 -\partial_y^2 \right)-\omega^2  & 0 & \frac{ik_xB_2}{\rho M_\mathrm{s}}  \\
\gamma B_2 \partial_y & \gamma i B_2 k_x  & 0 & \omega_\mathrm{my} &  -i\omega \\
0 & 0 & \gamma B_2 ik_x & i\omega& \omega_\mathrm{mz}
\end{bmatrix}\tilde{\bm{w}}_n(y) \equiv \bar{\varkappa}_\mathrm{mel}\cdot\tilde{\bm{w}}_n(y) = \begin{bmatrix} 0\\0\\0\\0\\0 \end{bmatrix},
\end{equation}
\end{widetext}
\noindent with the velocities $v_\parallel^2=C_{11}/\rho$, $v_\perp^2=C_{44}/\rho$, and $v_\asymp^2 = (C_{12}+C_{44})/\rho$, $C_{ij}$ the stiffness constants in Voigt notation and $\partial_y=\partial/\partial y$. Moreover,
\begin{eqnarray}
\omega_\mathrm{my} &=& \omega_0 + \omega_\mathrm{M} \left(\lambda_{\mathrm{ex}}k_\mathrm{tot}^2 + P\frac{\kappa_n^2}{k_\mathrm{tot}^2} \right) \,,\\
\omega_\mathrm{mz} &=& \omega_0+\omega_\mathrm{M} (\lambda_{\mathrm{ex}} k_\mathrm{tot}^2 +1-P)\,,\\
P &=& 1-\frac{1-e^{-k_\mathrm{tot}d}}{k_\mathrm{tot}d} \, ,
\end{eqnarray}
\noindent $k_\mathrm{tot}^2 = k_x^2+\kappa_n^2$, $\omega_0=\gamma_0 H_\mathrm{ext}$, $\omega_M=\gamma_0 M_\mathrm{s}$, $\lambda_{\mathrm{ex}}=\frac{2A_{\mathrm{ex}}}{\mu_0 M_\mathrm{s}^2}$, and $d$ the waveguide thickness. 

\begin{figure}[t]
	\includegraphics[width=7cm]{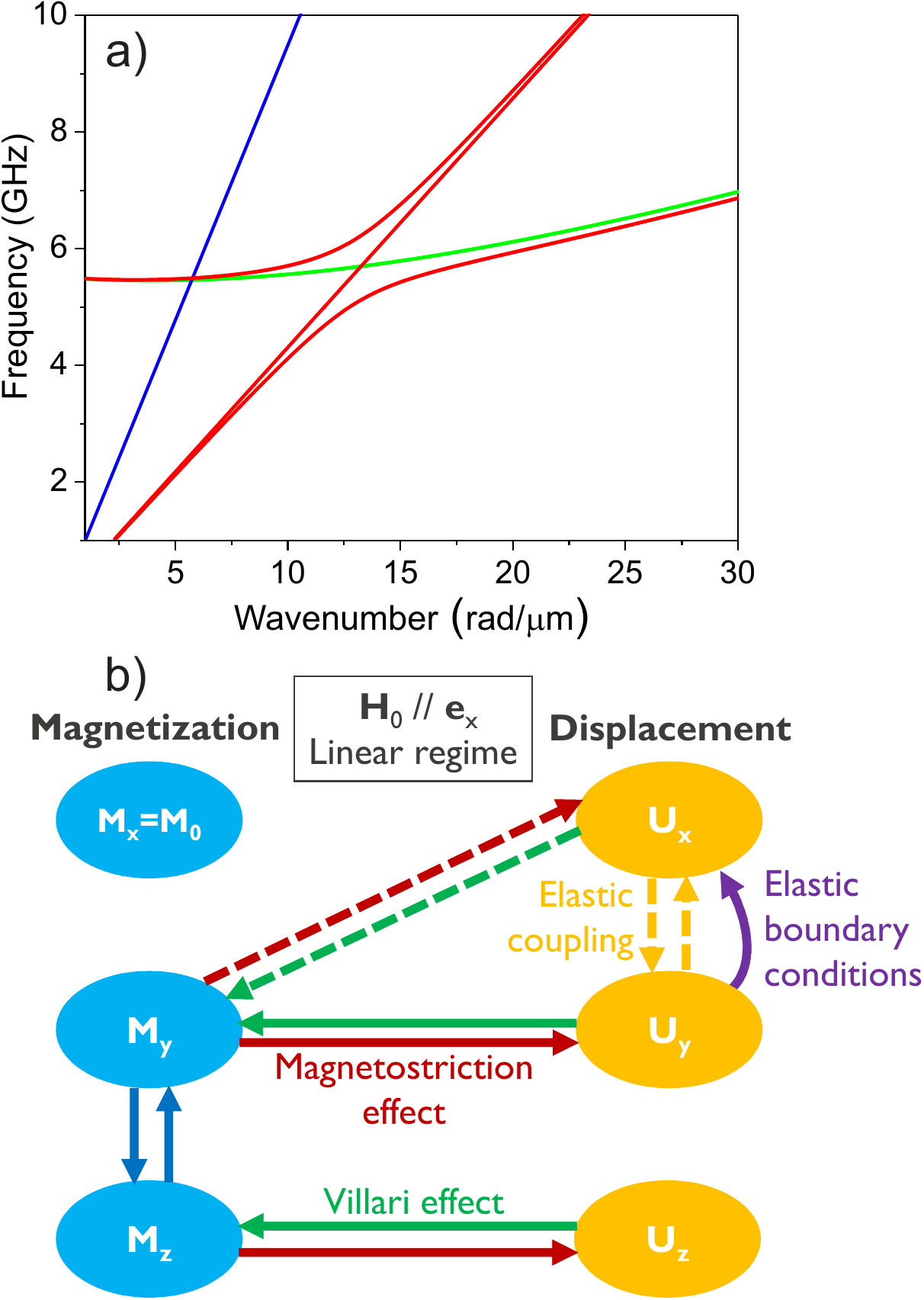}
	\caption{\textbf{(a)} Dispersion relations of magnetoelastic waves (red solid lines) in bulk CoFeB. The longitudinal elastic wave (blue) is uncoupled from the spin wave (green). The static external field has amplitude of $\mu_0H_0 = 25$ mT and is parallel to the static magnetization. \textbf{(b)} Schematic of interactions between the different dynamic components of a magnetoelastic wave. Solid arrows represent off-diagonal interaction terms in $\varkappa_\mathrm{mel}$ that are also present in bulk media or thin films. Dotted lines represent additional interactions that arise due to lateral confinement, as discussed in the text.} 
	\label{fig:scheme}
\end{figure}

Without magnetoelastic interactions, \emph{i.e.}\ for $B_1=B_2=0$, Eq.~\eqref{eq:lin_llg} leads to an eigensystem of purely elastic and magnetic (spin) waves in an isotropic waveguide, which are both well known \cite{Achenbach1973,Wang19,Demidov15}. For the geometry considered here, the dynamic in-plane displacement components represent laterally confined Lamb waves (LCLWs) whereas the out-of-plane displacement component corresponds to out-of-plane-polarized laterally confined shear waves, further called $P$ waves. For both wave types, an infinite amount of modes exists with either symmetric or antisymmetric mode profiles. For LCLWs, the symmetric ($S$) mode has symmetric $u_x$ and antisymmetric $u_y$ displacement profiles over the waveguide width, and \emph{vice versa} for antisymmetric ($A$) modes. In the magnetic system, the dynamic magnetization components represent confined backward volume spin waves (CBVSWs) as the static magnetic field is applied along the propagation direction. The lateral confinement of these spin waves also leads to modes with symmetric and antisymmetric profiles. For these modes, both $m_y$ and $m_z$ share the same symmetry with odd and even modes corresponding to symmetric and antisymmetric profiles, respectively. Note that this is in stark contrast to isotropic bulk systems, which only have a single spin wave mode in a given geometry (see Fig.~\ref{fig:scheme}a) \cite{Barber2004,Graff1975,Achenbach1973,Nayfeh1995}.

When magnetoelastic coupling is present, the confined elastic and spin waves mutually interact with each other. As discussed in more detail below, the magnetoelastic interaction, which is described by the off-diagonal terms in $\bar{\varkappa}_\mathrm{mel}$, results in an anticrossing of the dispersion relations of the elastic and magnetic waves \cite{Gurevich96,Tucker1972}. Near the anticrossing, the mutual interaction between the elastic and magnetic domain is strongest and can be quantified by the amplitude of the anticrossing gap, further called the magnetoelastic gap $\Delta f$. A larger magnetoelastic gap results in higher coupling rates and thus faster magnetoelastic energy oscillation between the magnetic and elastic domain during propagation \cite{Berk19}. Hence, $\Delta f$ is an important parameter for the description of magnetoelastic waves.  

For plane waves in an infinitely extended thin film, $\kappa_n$ as well as the partial derivatives $\partial_y$ are zero, and Eq.~\eqref{eq:lin_llg} reduces to a set of homogeneous linear equations. The system is solvable and allows for the derivation of an approximate analytical expression for the magnetoelastic gap based on the off-diagonal components of $\varkappa_\mathrm{mel}$ \cite{Gurevich96,Tucker1972}, which is given by
\begin{equation}
    2\pi\Delta f = \sqrt{\frac{2\gamma B_2^2 \omega_{mz}}{C_{44}M_s}}
    \label{MEL_gap_eq}
\end{equation}
\noindent More details about magnetoelastic waves in extended thin films can be found in Refs.~\cite{Kamenetskii21,Verba19}.

By contrast, the mode formation in a waveguide results in a spatial variation of $\tilde{\bm{w}}_n$ across the waveguide width (along $\bm{\hat{y}}$). This leads to two additional coupling terms between mechanical and magnetic components in $\varkappa_\mathrm{mel}$: (i) $\varkappa_{1,4}$ represents an additional mechanical body force originating from the mode profile of the magnetization components; and (ii) $\varkappa_{4,1}$ represents an action from the elastic on the magnetic system and stems from the mode profile of the longitudinal displacement component. Hence, the confinement also influences the magnetoelastic coupling itself, since both terms are absent for plane waves in bulk systems of thin films. Furthermore, the two additional terms depend on the shape of the mode profiles and thus every elastic or spin wave mode is expected to show a different magnetoelastic coupling behavior. 

The modification of the magnetoelastic coupling by lateral confinement has several consequences. First, in a waveguide, an infinite set of confined elastic and magnetic modes exist and interact with each other. As a result, numerous crossings of the dispersion relations exist, which may lead to the formation of magnetoelastic gaps at various frequencies and wavenumbers. By contrast, only a single magnetoelastic gap is formed in a bulk system, as shown in Fig. 2a. Moreover, the dispersion relation of the magnetoelastic waves as well as the magnetoelastic gap for the different modes is hard or even impossible to calculate analytically. Finally, the two additional components $\varkappa_{4,1}$ and $\varkappa_{1,4}$ result in interactions between the dynamic magnetization and the longitudinal displacement component. This is a specific effect of the confinement as the longitudinal displacement component is uncoupled from the magnetic system in bulk media (see Fig.~2a) or thin films \cite{Kamenetskii21, Gurevich96}. 

The magnetoelastic interactions between the different mechanical and magnetic components expressed by Eq.~\eqref{eq:lin_llg} are illustrated in Fig.~\ref{fig:scheme}b. Solid lines represent interaction terms, which are present both in bulk and laterally confined systems, whereas dotted lines represent additional interaction terms, which arise due to confinement and lateral mode formation. The coupling between magnetization and displacement is indicated by red and green lines: red lines correspond to the Villari effect that describes the change in magnetization due to strain (displacement gradient), whereas the green lines correspond to magnetostriction, which describes the change in displacement due to a change of the magnetization. 

Finally, we note that the $\varkappa_\mathrm{mel}$ tensor in Eq.~\eqref{eq:lin_llg} depends only on the $B_2$ coupling constant. Terms depending on $B_1$ are of second order in magnetization and displacement and can thus be neglected in the linear regime considered here. This means that shear strains couple much more strongly to the magnetization than normal strains. This conclusion is analogous to the results of spin wave excitation by local magnetoelastic transducers \cite{Duflou17}. 

\section{Numerical simulations of magnetoelastic waves in thin waveguides}
\label{sec:sim_results}

\subsection{Numerical approach}

As discussed above, it is possible to find analytical solutions of the two coupled differential equations of motion \eqref{eq:llg} and \eqref{eq:elastodynamic} for bulk media and thin films, as they can be reduced to a homogeneous set of linear equations and therefore to an eigenvalue problem. However, for thin waveguides, this is not possible, and a set of coupled partial differential equations remains. Even for simple geometries, such as a linear thin waveguide, it is therefore more practicable to solve the equations numerically. This is even more the case for complex geometries (and more complex boundary conditions), which render analytical solutions impossible.

For this purpose, we have extended the micromagnetic software package mumax3 \cite{Vansteenkiste14} by complementing the already implemented LLG equation \eqref{eq:llg} and magnetic boundary conditions Eq.~\eqref{eq:M_boundary} with Eqs.~\eqref{eq:E_mel} to \eqref{eq:E_kin} and \eqref{eq:elastodynamic} to \eqref{eq:U_boundary}, which allows for the simulation of magnetoelastic waves in arbitrary geometries \cite{github}. The extension is based on a finite difference approach to simultaneously solve the magneto- and elastodynamic differential equations. Several different solver algorithms have been implemented such as the Euler, Heun, fourth-order Runge-Kutta (RK4), and leapfrog methods. All methods gave essentially identical results. However, the RK4 method provided the best performance in most cases and was therefore used for the simulations below. Furthermore, all mathematical operations have been implemented on GPUs, which strongly reduces the computation time \cite{Leliaert_2018}.

\subsection{Simulation details}

The simulated system is schematically represented in Fig.~\ref{fig:geometry} and consists of a thin nanoscale CoFeB waveguide with a thickness of $d=20$~nm, a width of $w=200$~nm, and a length of $\ell=10$~$\mu$m. The mesh cell size was set to $5\times 5\times 20$ nm$^3$, which is much smaller than the wavelength of the studied magnetoelastic waves and of the same order as the magnetic exchange length of CoFeB ($\sim 4.5$~nm). Along $\bm{\hat{z}}$, the waveguide was modeled by a single cell as the dynamic displacement and magnetization components are approximately uniform over the thickness. 

The material parameters of CoFeB were extracted from the literature: a saturation magnetization of $M_\mathrm{s}=1.2$~MA/m \cite{Yu15}, an exchange constant of $A_\mathrm{ex}=18$ pJ/m \cite{Conca14}, a Gilbert damping constant of $\alpha=0.004$, a mass density of $\rho=8$~kg/m$^3$ \cite{Peng16},  magnetoelastic coupling constants of $B_1=B_2=-8.8$~MJ/m$^3$ \cite{Gueye16}, as well as the stiffness constants $C_{11}=283$~GPa, $C_{12}=166$~GPa, and $C_{44}=58$~GPa \cite{Gueye16}. The elastic damping was neglected in the simulations and therefore $\eta=0$. We remark that nonzero elastic damping mainly leads to a line broadening with weak expected effects on the dispersion relations and mode profiles. At both ends of the waveguide, the elastic and magnetic damping increased exponentially to $\alpha=0.5$ and $\eta = 5\times 10^{13}$ Ns/m$^4$ over a 1~$\mu$m long region to avoid reflection of the waves. A static external magnetic field of $\mu_0H_\mathrm{ext}=5$~mT was applied along $\bm{\hat{x}}$, \emph{i.e.} along the waveguide. Together with the demagnetization field due to the shape anisotropy, this was enough to saturate the magnetization along the waveguide without magnetization nonuniformities at the ends of the waveguide.

The magnetoelastic waves were magnetically excited by applying an rectangular 20 ps long magnetic field pulse in the center of the waveguide, as shown in Fig.~\ref{fig:geometry}. The excitation region spanned the full waveguide width and had a length of 100~nm. The amplitude of the excitation field pulse was $\mu_0h_\mathrm{ex}=1$~mT and the total duration of the simulations was 10 ns. Note that the calculated magnetoelastic gaps were proportional to $B_2$, in keeping with Eq.~\eqref{MEL_gap_eq}, indicating that the simulations were carried out in the linear magnetoelastic coupling regime.

\subsection{Dispersion relations and mode profiles of confined magnetoelastic waves}

\begin{figure}[t]
	\includegraphics[width=7.5cm]{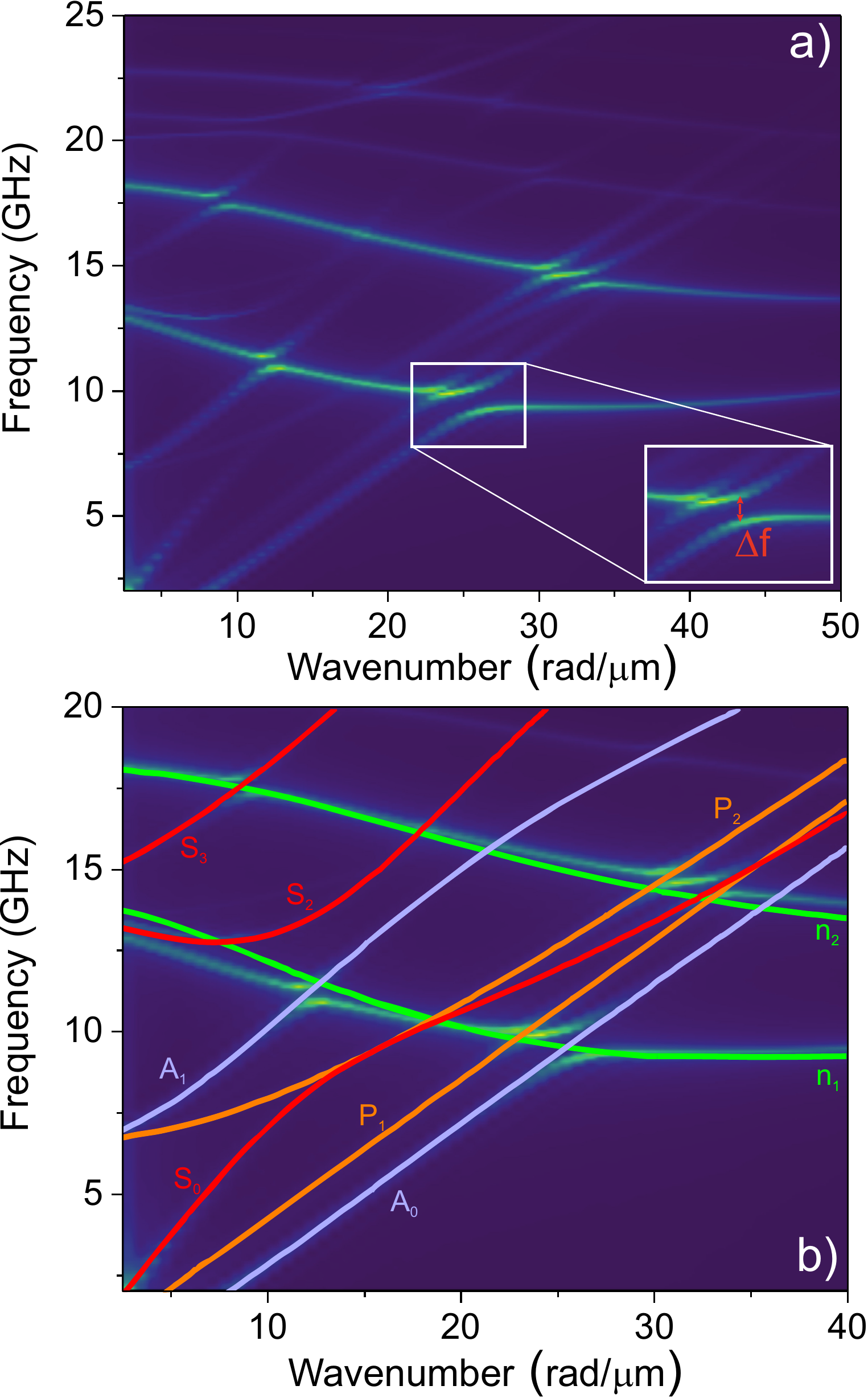}
	\caption{Dispersion relations of magnetoelastic waves in a 200 nm wide and 20 nm thick CoFeB waveguide. \textbf{(a)} Numerically calculated dispersion relations. The inset shows a magnification of a region near a magnetoelastic gap. \textbf{(b)} Analytically calculated dispersion relations of uncoupled elastic and spin waves superimposed to the numerically obtained results.}
	\label{fig:mode_coupling}
\end{figure}

The dispersion relations of confined magnetoelastic waves in the CoFeB waveguide were obtained by two-dimensional temporal and spatial (along $\bm{\hat{x}}$) Fourier transforms of the different magnetization and displacement components after pulsed excitation and are depicted in Fig.~\ref{fig:mode_coupling}a. To identify the different branches, the dispersion relations of the confined elastic and magnetic waves \textit{without} magnetoelastic interaction were also analytically calculated and are plotted over the numerically obtained magnetoelastic dispersion relations in Fig.~\ref{fig:mode_coupling}b. Here, the green solid lines correspond to the first two CBVSW width modes, given by $\omega_n = \sqrt{\omega_{my}\omega_{mz}}$ \cite{Kalinikos86}. By contrast, blue and red solid lines correspond to $A$- and $S$-type LCLWs, respectively, which are described by the solutions of \cite{Achenbach1973}
\begin{equation}
\label{eq:disp_lamb}
\frac{\tan\left(bd\right)}{\tan\left(ad\right)} = - \left[\frac{4k_x^2ab}{\left(k_x^2-b^2\right)^2}\right]^{\pm1},
\end{equation}
\noindent with $a^2 = \left(\frac{\omega}{v_\parallel}\right)^2 + k_x^2$ and $b^2 = \left(\frac{\omega}{v_\perp}\right)^2 + k_x^2$. Here, the plus sign in the exponent describes $S$ modes, whereas the minus sign describes $A$ modes. Finally, orange solid lines correspond to $P$ waves, whose dispersion relations are described by \cite{Achenbach1973}
\begin{equation}
\label{eq:disp_uz}
\omega_m^2 = v_\perp \left[k_x^2 + \left(\frac{(m-1)\pi}{w}\right)^2\right],
\end{equation}
\noindent with $w$ the waveguide width and $m$ the mode number.

In Fig.~\ref{fig:mode_coupling}, three different regimes can be identified. Far from crossover points, the numerically calculated dispersion relations of confined magnetoelastic waves coincide closely with the analytical dispersion relations of uncoupled confined elastic and magnetic modes. Small differences between analytical CBVSW and numerical magnetoelastic dispersion relations stem rather from the finite waveguide size, in which the wavelength can become comparable to the waveguide length and the analytical treatment becomes less accurate \cite{Wang19}. In addition, the data show that a multitude of magnetoelastic gap regions exist when elastic and magnetic wave dispersions cross. Note that the number of crossings is much larger than \emph{e.g.} for bulk systems (see Fig.~\ref{fig:scheme}a) or thin films \cite{Kamenetskii21,Gurevich96} due to the large number of distinct confined elastic and magnetic modes.

We first discuss regions far from the interaction points, where the dispersion relation can be considered as quasi-elastic or quasi-magnetic. They occur when the intrinsic elastic and magnetic resonance frequencies are strongly mismatched for a given wavenumber. As discussed in more detail below, the energy of quasi-elastic or quasi-magnetic waves is then almost completely transported in the elastic or magnetic domains, respectively \cite{Gurevich96,Tucker1972}.

\begin{figure*}[t]
	\includegraphics[width=16cm]{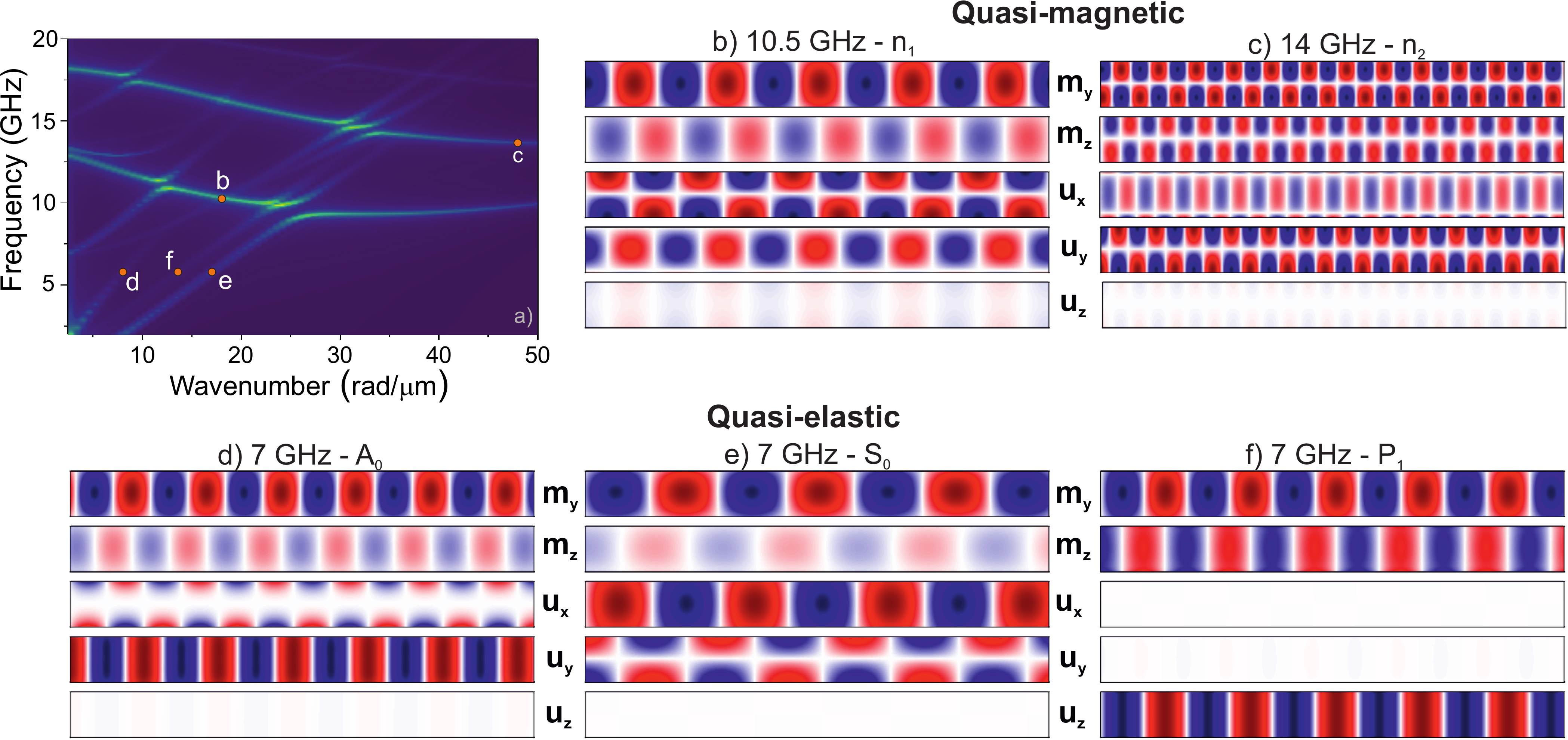}
	\caption{Profiles of  magnetoelastic wave components in the ``quasi'' regimes. \textbf{(a)} Dispersion relation of magnetoelastic waves indicating the frequencies and wavenumbers of the different modes depicted in (b)--(f). Snapshots of displacement and magnetic component profiles of $n_1$ and $n_2$ quasi-magnetic waves are shown in \textbf{(b)} and \textbf{(c)} for frequencies of 10.5 and 14.0 GHz (wavenumbers of 17 rad/$\mu$m and 46 rad/$\mu$m), respectively. \textbf{(d)}, \textbf{(e)}, and \textbf{(f)} show snapshots of displacement and magnetic component profile of $A_0$, $S_0$, and $P_1$ quasi-elastic waves at 7.0 GHz and wavenumbers of 9 rad/$\mu$m, 15 rad/$\mu$m, and 19 rad/$\mu$m, respectively.}
	\label{fig:mode_profile1}
\end{figure*}

The mode profiles of confined quasi-magnetic and quasi-elastic waves are shown in Fig.~\ref{fig:mode_profile1} for selected points in the dispersion relations (Fig.~\ref{fig:mode_profile1}a). Figures~\ref{fig:mode_profile1}b and \ref{fig:mode_profile1}c depict snapshot images of the magnetization and displacement components of two quasi-magnetic waves at frequencies of 10.5 GHz and 14.0 GHz, respectively. The magnetization dynamics are characterized by wave-like $m_y$ and $m_z$ components with a relative phase shift of $\pi$. The $m_z$ component is weaker than $m_y$ due to the ellipticity of the magnetization precession \cite{Gurevich96}. At 10.5 GHz, a single amplitude maximum is found across the waveguide, as expected for a symmetric first-order width mode ($n_1$). By contrast, the mode at 14.0 GHz shows two amplitude maxima and is therefore an antisymmetric second-order width mode ($n_2$). These mode profiles are essentially identical to those of uncoupled CBVSWs obtained for $B_1 = B_2 = 0$, which shows that the presence of the magnetoelastic coupling does not affect CBVSWs  in the quasi-magnetic regime. 

Nonetheless, quasi-magnetic waves also possess accompanying elastic waves, as shown in Figs.~\ref{fig:mode_profile1}b and \ref{fig:mode_profile1}c. Note that a symmetric (antisymmetric) spin wave mode leads to symmetric (antisymmetric) $u_y$ and $u_z$ components, as well as to an antisymmetric (symmetric) $u_x$ component. By contrast, the $u_z$ component is typically very weak. The impact of the symmetry on the coupling of the different components will be discussed further below.

Figures~\ref{fig:mode_profile1}d to \ref{fig:mode_profile1}f represent snapshot images of the magnetization and displacement components of three quasi-elastic waves at 7.0 GHz and wavenumbers as shown in Fig.~\ref{fig:mode_profile1}a. The displacement components of the first mode (Fig.~\ref{fig:mode_profile1}d) correspond to those of an $A$-type LCLW, whereas the second mode (Fig.~\ref{fig:mode_profile1}e) corresponds to an $S$-type LCLW. Finally, the third mode (Fig.~\ref{fig:mode_profile1}f) can be linked to an elastic $P$ wave. In all cases, the mode displacement patterns are not significantly affected by the magnetoelastic interaction. The accompanying CBVSW modes are symmetric ($n_1$) in all types of quasi-elastic waves. Again, this will be discussed in further detail below. 

We now turn to regions in reciprocal space where the dispersion relations of confined spin waves and elastic waves intersect. In these regions, strong magnetoelastic interactions lead to an anticrossing behavior and the formation of a gap in the dispersion relation. The resulting waves are confined magnetoelastic waves and the transported energy oscillates between the elastic and magnetic domains during propagation \cite{Gurevich96,Tucker1972}. The magnetoelastic gap $\Delta f$ quantifies the interaction strength between the different constituting modes. A detailed look a Fig.~\ref{fig:mode_coupling}a reveals that the magnitude of $\Delta f$ strongly varies for the different anticrossing points, which means that the coupling dependents on elastic and spin wave modes. This can be related to the symmetry and the spatial profiles of the magnetic and elastic waves as well as the resulting interaction terms and will be discussed in more detail in the next section.

\begin{figure}[t]
	\includegraphics[width=7.5cm]{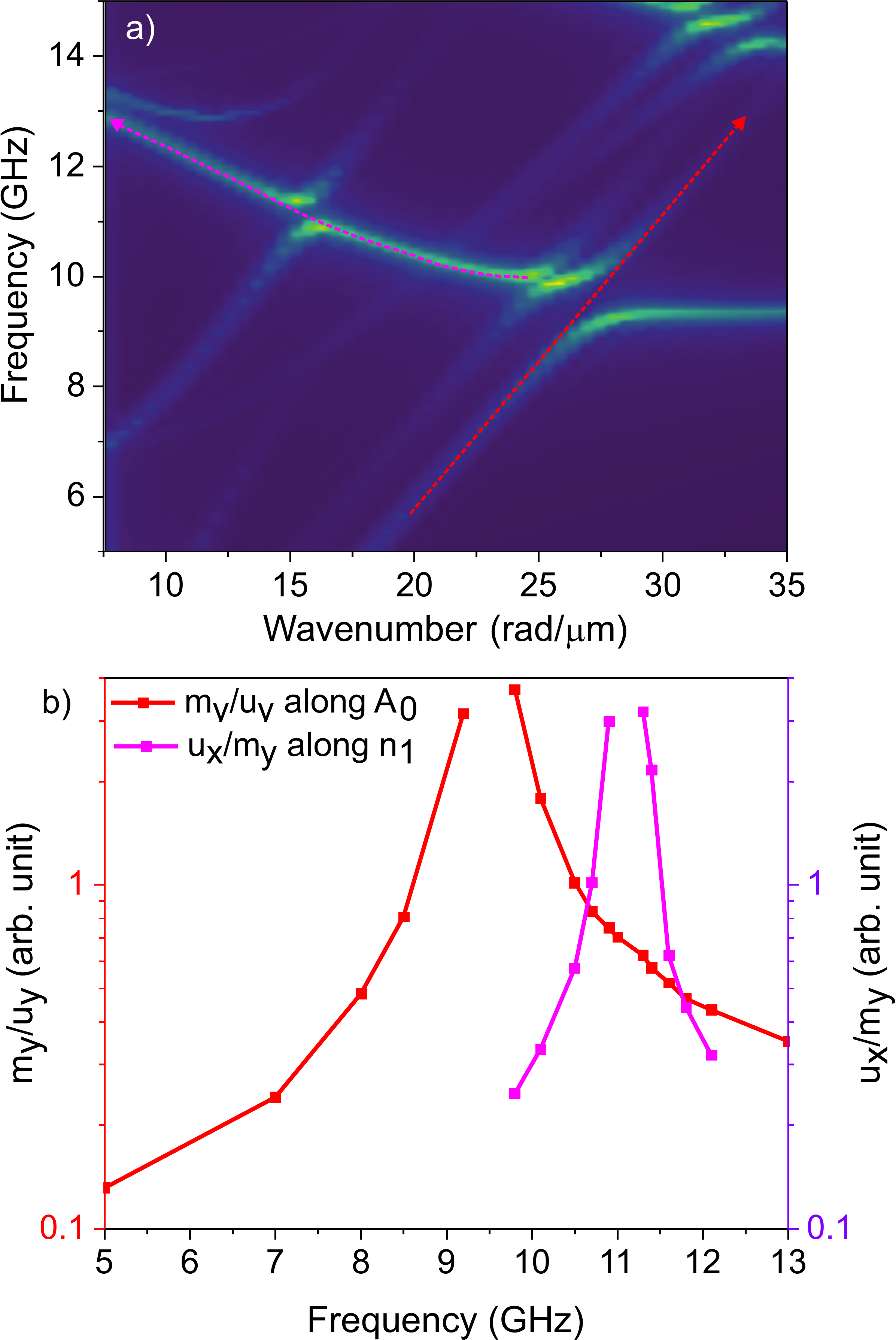}
	\caption{Amplitude ratio of the displacement and magnetization components of magnetoelastic waves along acoustic-like quasi-elastic $A_0$ (red dashed line) and spin-wave-like quasi-magnetic $n_1$ (pink dashed line) branches of the dispersion relation. \textbf{(a)} shows the trajectories in frequency-wavenumber space. \textbf{(b)} $m_y/u_y$ ratio along the $A_0$ dispersion branch (solid red line), as well as $u_x/m_y$ ratio along the quasi-magnetic $n_1$ (solid pink line) dispersion branches. Near the magnetoelastic gap, the ratios peak and decay rapidly further away.}
	\label{fig:ellipticity}
\end{figure}

The transition between magnetoelastic and quasi-static or quasi-magnetic waves is illustrated in Fig.~\ref{fig:ellipticity}, which shows the ratio of the magnetic ($m_x$, $m_y$) and elastic ($u_x$,$u_y$) components of magnetoelastic waves along an acoustic-like (pink dashed line in Fig.~\ref{fig:ellipticity}a) and spin-wave-like (red dashed line in Fig.~\ref{fig:ellipticity}a) branch of the dispersion relation. The data in Fig.~\ref{fig:ellipticity}b show that the relative amplitude of magnetic (elastic) components along the acoustic (spin-wave) branch of the dispersion relation decreases rapidly away from the magnetoelastic gap. A strong decrease is already seen a few 100 MHz away from the gap. Further away, the relative intensity becomes low and the energy is mainly transported in the elastic (magnetic) domain. In this case, the waves can be considered as quasi-elastic (quasi-magnetic).

\begin{figure}[bt]
	\includegraphics[width=9cm]{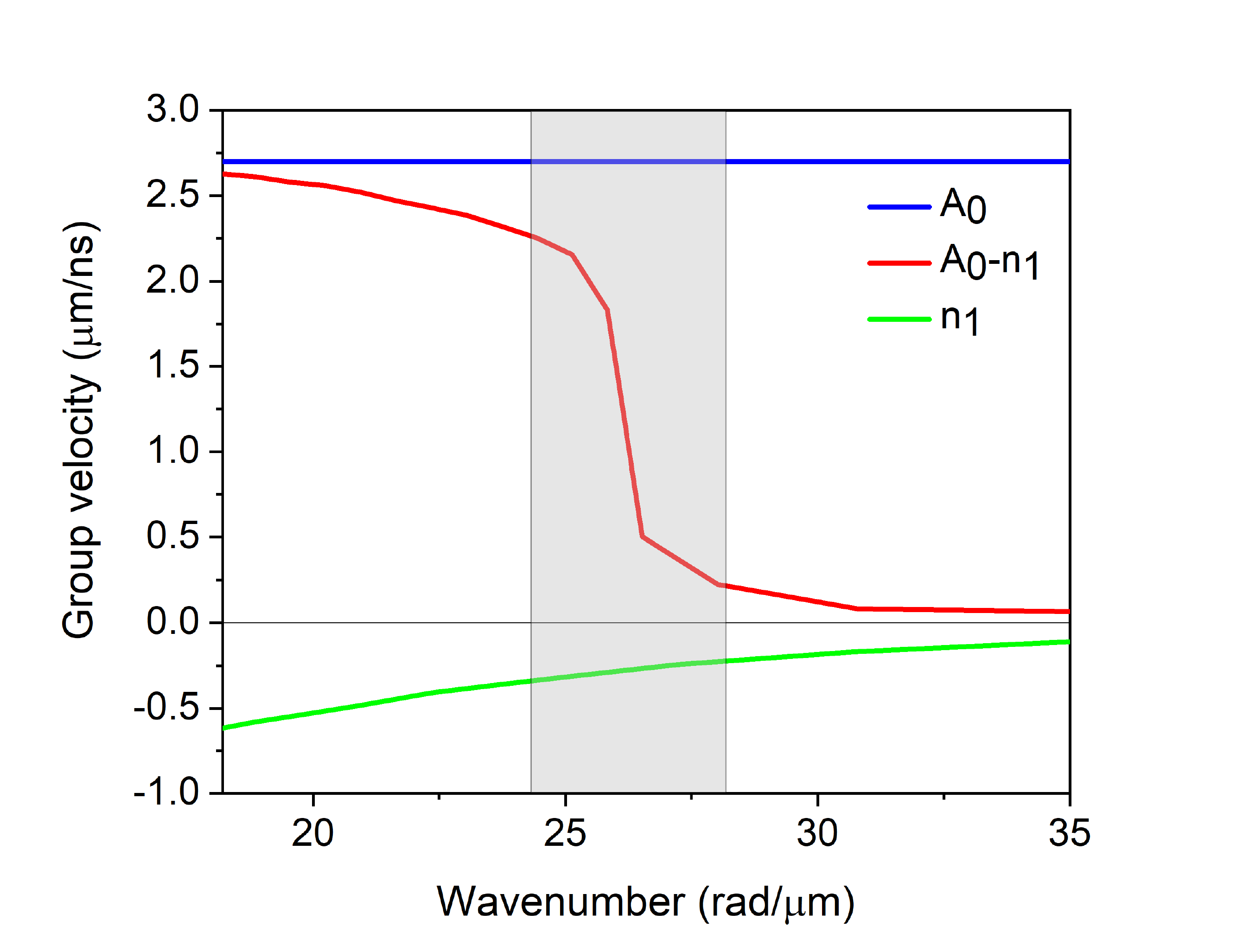}
	\caption{Group velocities as a function of wavenumber for the $A_0$ LCLW (blue solid line), the $n_1$ CBVSW (green solid line) and the coupled $n_1$--$A_0$ magnetoelastic wave mode (red solid line). The grey area represents the anticrossing region.}
	\label{fig:vg}
\end{figure}

Near the anticrossing points, the dispersion relation of magnetoelastic wave differs strongly from those of uncoupled elastic or magnetic waves, which also affects the group velocity $\bm{v}_\mathrm{g}=\partial \omega/\partial \bm{k}$. As an illustrative example, Fig.~\ref{fig:vg} represents the calculated group velocity near the $n_1$--$A_0$ intersection (see Fig.~\ref{fig:mode_coupling}a), where the $n_1$ CBVSW and $A_0$ LCLW modes cross. The blue and green solid lines correspond to uncoupled $A$-type LCLWs and CBVSWs, respectively. The CBVSW is characterized by a small negative group velocity, as typical for spin waves in this geometry with the magnetization parallel to the wavevector $k$ in the dipolar regime. By contrast, the group velocity of the magnetoelastic wave (red solid line) is strongly modified near the gap (grey region). At small wavenumbers, the group velocity approximates the group velocity of the elastic $A_0$ LCLW mode, whereas at higher wavenumbers, the group velocity converges to the CBVSW group velocity. This shows that magnetoelastic coupling can strongly increase the group velocity of the waves in a specific frequency or wavenumber range. Moreover, the data shows that even the sign of the group velocity can be changed by magnetoelastic interactions far from the magnetoelastic gap, which indicates that magnetoelastic interactions can be crucial for tuning the properties of CBVSWs in scaled waveguides. 

\begin{figure*}[t]
	\includegraphics[width=16cm]{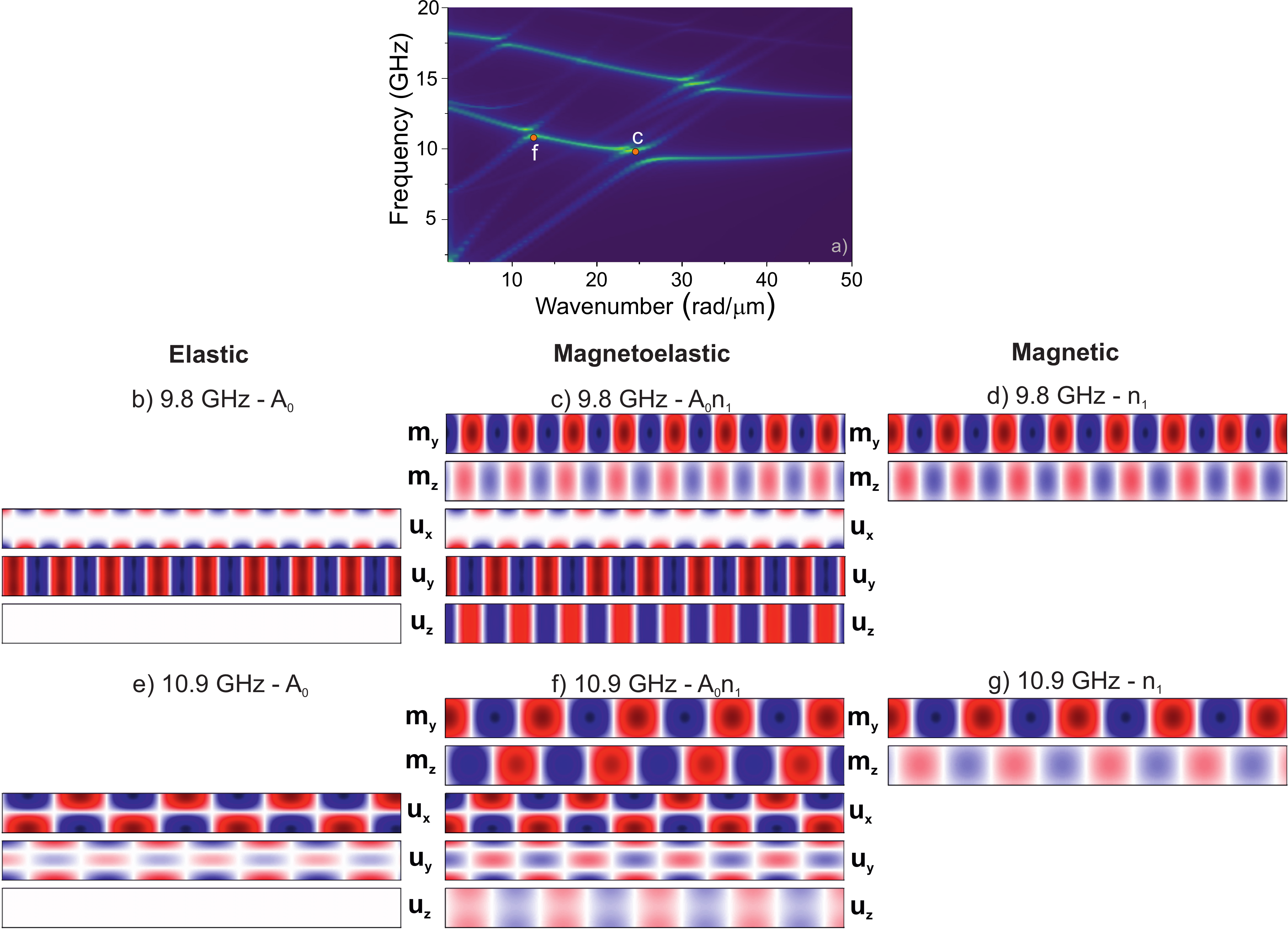}
	\caption{Profiles of magnetoelastic wave components in the strongly coupled magnetoelastic regime. \textbf{(a)} Dispersion relation of magnetoelastic waves indicating the frequencies and wavenumbers of coupled $n_1-A_0$ and $n_1-A_1$ modes. \textbf{(c)} and \textbf{(f)} show snapshots of displacement and magnetization profiles of $n_1-A_0$ and $n_1-A_1$ magnetoelastic waves at frequencies of 9.8 and 10.9 GHz (wavenumbers of 15 rad/$\mu$m and 12 rad/$\mu$m), respectively. For comparison, the mode profiles of uncoupled elastic [\textbf{(b)} and \textbf{(c)}] and magnetic [\textbf{(d)} and \textbf{(g)}] waves are also shown.}
	\label{fig:mode_profile2}
\end{figure*} 

We finally discuss the mode profiles of magnetoelastic waves in the magnetoelastic gap region. Figure~\ref{fig:mode_profile2} shows snapshot images of the magnetization and displacement components of two magnetolastic waves near the $n_1$--$A_1$ and $n_1$--$A_0$ intersections. For comparison, mode profiles are also shown for uncoupled LCLWs (left side) and CBVSWs (right side). While the mode profiles themselves change with frequency and mode number, the data indicate that they are not qualitatively modified by the magnetoelastic interactions and the overall displacement and magnetization modes remain similar. However, two profile modifications can be observed: (i) the magnetoelastic coupling leads to a large amplitude of the $u_z$ component, which is not present in uncoupled LCLWs; (ii) the ellipticity of the magnetization precession is reduced, since the relative intensity of the $m_z$ component with respect to the $m_y$ component is increased, especially for the $n_1$--$A_0$ mode at 10.9 GHz. 

\subsection{Symmetry of magnetic and elastic mode profiles and the impact on the magnetoelastic coupling}

In the previous section, we have discussed dispersion relations and mode profiles of confined magnetoelastic waves in narrow CoFeB waveguides. A closer look at the dispersion relations in Fig.~\ref{fig:mode_coupling} reveals that the magnitude of the magnetoelastic gap strongly depends on the interacting elastic and magnetic waves. To shed light on the influence of the mode profiles of uncoupled waves on the magnetoelastic interaction, the system can be decomposed into magnetic and elastic subsystems, which interact via the magnetoelastic field $\bm{h}_\mathrm{mel}$ and the magnetoelastic force $\bm{f}_\mathrm{mel}$. The magnetoelastic force $\bm{f}_\mathrm{mel}$ originates from the spatial variation of the magnetization as described by Eq.~\eqref{eq:Fmel_guide} in App.~A. In the linear regime and for a waveguide geometry, it is given by
\begin{equation}
\label{eq:Fmagnetoelastic_guide_lin}
\mathbf{f}_\mathrm{mel} =  B_2\begin{bmatrix}
 \frac{\partial m_y}{\partial y}   \\ 
 ik_x m_y   \\
 ik_xm_z  \end{bmatrix} \,.
\end{equation}
Hence, $f_x \propto \partial m_y / \partial y$ and $f_{y,z} \propto m_{y,z}$ with $m_{y,z}$ of the form in Eq.~\eqref{eq:mode_profile}. The mode profiles are characterized by trigonometric functions and thus derivatives of even mode profiles lead to odd mode profiles and \emph{vice versa}. Moreover, a symmetric (antisymmetric) $f_x$ component is complemented by antisymmetric (symmetric) $f_y$ and $f_z$ components, which share the symmetry of the dynamic magnetization. 

\begin{figure*}[tb]
	\includegraphics[width=16cm]{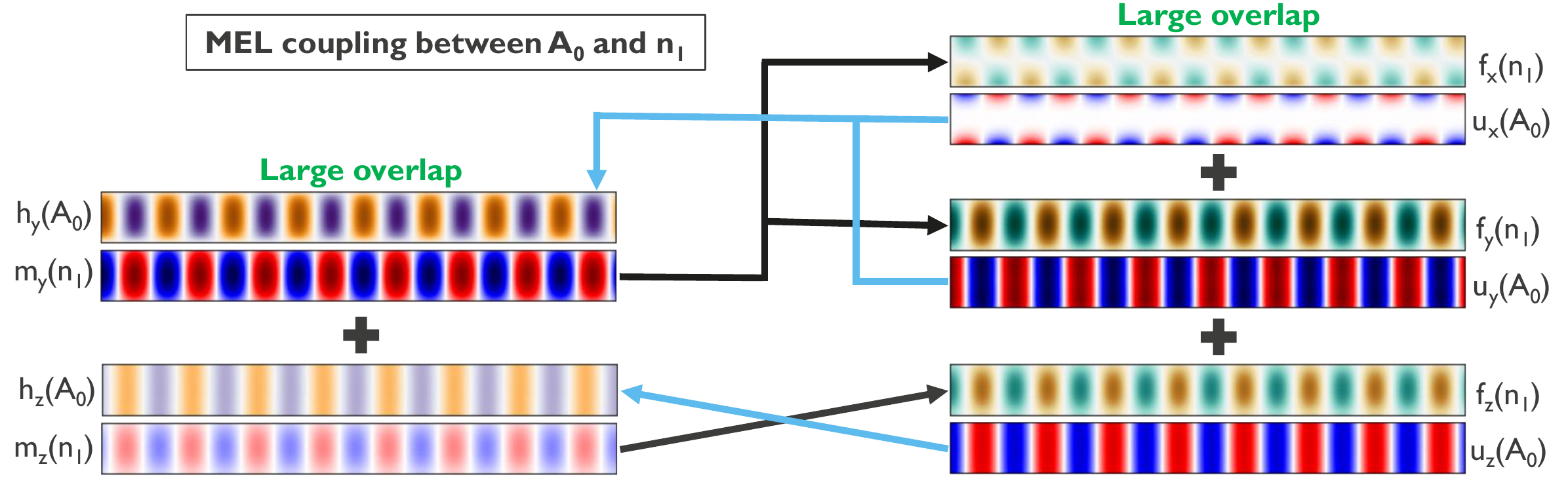}
	\caption{Graphical explanation of the coupling between the magnetic $n_1$ and elastic $A_0$ modes. The profiles of the magnetization and magnetoelastic force corresponding to the magnetic $n_1$ mode are plotted as well as the profiles of the displacement and magnetoelastic field corresponding to the elastic $A_0$ mode.} 
	\label{fig:overlaps}
\end{figure*}

\begin{figure*}[tb]
	\includegraphics[width=16cm]{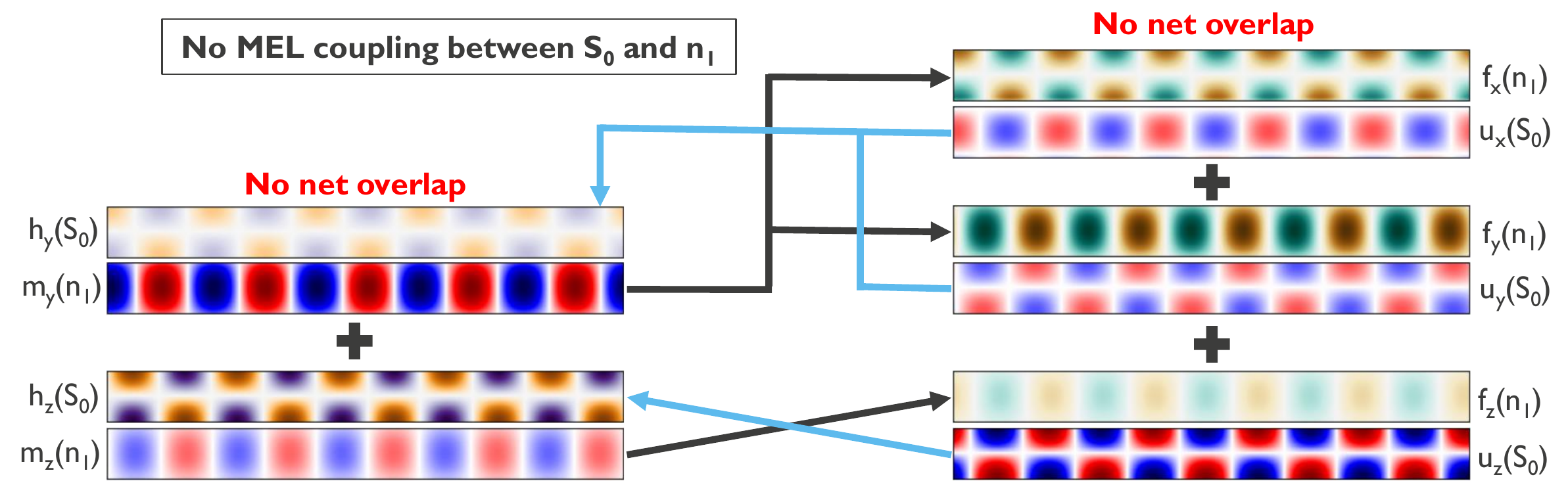}
	\caption{Graphical explanation of the coupling between the magnetic $n_1$ and elastic $S_0$ modes. The profiles of the magnetization and magnetoelastic force corresponding to the magnetic $n_1$ mode are plotted as well as the profiles of the displacement and magnetoelastic field corresponding to the elastic $S_0$ mode.} 
	\label{fig:overlaps2}
\end{figure*}

The magnitude of the elastodynamics generated by a mechanical force is proportional to the overlap integral between the resonant displacement mode profile and the applied force. Hence, the generation efficiency of the $j$\textsuperscript{th} elastic mode by the $i$\textsuperscript{th} spin wave mode is given by
\begin{equation}
\label{eq:u_overlap}
\xi_{i,j}^{\mathrm{mag \to el}} \propto \left| \int_V  \bm{u}_j^* \cdot \bm{f}_{\mathrm{mel},i} dV\right| \, ,
\end{equation}
\noindent where the asterisk denotes the complex conjugate and $V$ the volume of the system. As discussed in Sec. \ref{sec:theory}, $A$-type LCLWs possess antisymmetric $u_x$ and symmetric $u_y$ components, whereas $S$-type LCLWs  possess symmetric $u_x$ and antisymmetric $u_y$ components. Hence, odd magnetic modes with symmetric magnetization profiles entail magnetoelastic forces that strongly overlap with $A$-type LCLWs. This is illustrated in Fig.~\ref{fig:overlaps}, which shows profiles of the displacement components of an $A$-type LCLW as well as the magnetoelastic force $\bm{f}_\mathrm{mel}$ generated by an $n_1$ CBVSW. The corresponding profiles for $S$-type LCLWs are shown in Fig.~\ref{fig:overlaps2}.

Conversely, the elastodynamics generate a magnetoelastic field that interacts with the magnetization. According to Eq.~\eqref{eq:hmel_guide}, the magnetoelastic field for a waveguide geometry in the linear regime is given by
\begin{equation}
\label{eq:hmagnetoelastic_guide_lin}
\mathbf{h}_{\mathrm{mel}} =  -\frac{1}{\mu_0M_\mathrm{s}}  \begin{bmatrix}
2B_1ik_xu_x  \\
B_2\left(\frac{\partial u_x}{\partial y}+ik_xu_y \right) \\
B_2ik_xu_z 
\end{bmatrix}\,.
\end{equation}
\noindent For a waveguide magnetized along $\bm{\hat{x}}$, $h_x$ has no influence on the magnetization dynamics as it generates no magnetic torque. By contrast, the $h_y$ and $h_z$ components exert a torque on the magnetization and depend on the displacement profiles (see Figs.~\ref{fig:overlaps} and \ref{fig:overlaps2} for $A$-type and $S$-type LCLWs, respectively). More specifically, the $h_y$ component depends on both the $u_x$ and $u_y$ components, whereas the $h_z$ component only depends on $u_z$. This indicates that LCLWs and $P$ waves couple differently to the magnetization. Whereas LCLWs generate a $h_y$ component, $P$ waves couple through the $h_z$ component. The coupling to LCLWs is complicated by the two terms in $h_y$, which are proportional to $\partial u_x / \partial y$ and $u_y$, respectively. For a symmetric (antisymmetric) $u_x$ displacement profile, the derivative is antisymmetric (symmetric) and therefore shares the symmetry of the $u_y$ component (see Sec. \ref{sec:theory2}). Hence, $h_y$ possesses the same symmetry as $u_y$ for a LCLW. 

The same above symmetry considerations hold thus for the magnetoelastic force excitation of elastodynamics. The excitation efficiency of the $i$\textsuperscript{th} spin wave mode by the $j$\textsuperscript{th} elastic mode is proportional to 
\begin{equation}
\label{eq:m_overlap}
\xi_{j,i}^{\mathrm{el \to mag}} \propto \left| \int_V  \bm{m}_i^*\cdot\bm{h}_{\mathrm{mel},j} dV\right|\,.
\end{equation}
\noindent Again, $h_x$ does not contribute to the dynamics as the dynamic magnetization along $\bm{\hat{x}}$ is zero. Hence, Eqs.~\eqref{eq:hmagnetoelastic_guide_lin} and \eqref{eq:m_overlap} indicate that $A$-type LCLWs couple to antisymmetric (even) CBVSW modes, whereas $S$-type LCLWs couple to symmetric (odd) CBVSW modes. By contrast, elastic $P$ waves interact with CBVSW modes with the same symmetry and mode number as the $P$ waves themselves. 

Combining both elastic and magnetic subsystems demonstrates that there are strong mutual interactions between pairs of elastic and magnetic waves of a given symmetry: (i) $S$-type LCLWs and odd CBVSW modes; (ii) $A$-type LCLWs and even CBVSW modes; as well as (iii) elastic $P$ wave modes and CBVSW modes with the same symmetry. The symmetry considerations explain well the numerical results in Fig.~\ref{fig:mode_coupling}, where a high magnetoelastic gap indicates strong interaction. 

It is interesting to compare the calculated magnitudes of the magnetoelastic gaps for waveguides to bulk systems or thin films. The strongest coupling (largest gap) for the studied CoFeB waveguide is observed for the interaction of the $A_0$ LCLW and $n_1$ CBVSW modes with a gap of $\Delta f = 0.9$~GHz. By contrast, the magnetoelastic gap for a bulk system with identical CoFeB material parameters is $\Delta f = 1.0$~GHz (\emph{cf.}~Fig.~\ref{fig:scheme}a). Furthermore, the magnetoelastic gap for higher order modes (both elastic and spin wave) consistently decreases with increasing mode numbers. These results suggest that the magnetoelastic interaction is reduced by confinement in waveguide structures and that the bulk gap represents an upper limit for magnetoelastic gaps in confined systems.

\section{Conclusions}

In conclusion, we have presented a combined analytical and numerical study of confined magnetoelastic waves in a nanoscale CoFeB waveguide with the static magnetization parallel to the propagation direction. The equations of motion for magnetoelastic waves in this waveguide geometry differ from those for bulk or thin film systems due to the lateral nonuniformity of the mode profiles of uncoupled elastic and spin waves. As a result, additional coupling terms appear, and the magnetic system also couples to longitudinal displacement components, unlike for bulk systems. In addition, the linearized differential equations indicate that only the $B_2$ coupling constant is of importance in the linear regime. This means that only shear strains affect the coupling whereas the influence of normal strains is much weaker.

The equations of motions for confined magnetoelastic waves consist of a system of partial differential equations, which cannot be solved analytically, especially for complex geometries. To gain more insight in the behavior of confined magnetoelastic waves in waveguides, the micromagnetic solver mumax3 was extended to include elastodynamics as well as the magnetoelastic coupling. This approach allowed for the calculation of the dispersion relations of magnetoelastic waves in the CoFeB waveguide. The numerical results demonstrate that the mode-dependent magnetoelastic coupling can be understood by the mode profile symmetry of uncoupled confined elastic and magnetic waves in the waveguide. In addition, it was found that the mode profiles of the elastic and magnetic components are only weakly affected by the magnetoelastic coupling. 

The results further show that the group velocities of confined magnetoelastic waves can be much larger than those of uncoupled CBVSWs and even be of the opposite sign. Moreover, the numerical procedure also allowed for the analysis of the eigenstates of the system. The results indicate a very strong decay of the dynamic magnetization from the magnetoelastic gap regime towards the quasi-elastic regime. An analogous behavior was found for the elastic displacement, which decreases rapidly from the magnetoelastic gap towards the quasi-magnetic regime.

This work opens new perspectives for the usage of magnetoelastic waves for spintronic applications, in which information is transported via waves in waveguides \cite{Khitun11,Talmelli19,Fischer17,Mahmoud20}. Considering an isolated waveguide, approaches to couple mechanical degrees of freedom to spin waves based on (inverse) magnetostriction are expected to generate magnetoelastic waves rather than pure spin waves in magnetostrictive waveguides. The above results demonstrate differences as well as similarities of the magnetoelastic and noninteraction systems that can provide a better understanding of the underlying wave properties in future magnonic experiments. Moreover, the numerical approach by extending the mumax3 micromagnetic solver paves the way for the more detailed studies of confined magnetoelastic waves, including of their nonlinear dynamics.

\appendix
\section{Linearized equations of motion for a waveguide geometry}
\label{app:A}

In this appendix, linearized equations of motion are derived for magnetoelastic waves propagating in thin magnetostrictive waveguides with the static magnetization along their long axis $\bm{\hat{x}}$. Starting with general expressions of effective magnetic fields and mechanical body forces for a waveguide geometry, the equations of motion are subsequently derived. An ansatz of a propagating wave along the waveguide leads to the coupled homogeneous system of partial differential equations in Eq.~\eqref{eq:system_of_melwaves}, which describes the eigenmodes of the system.

The constitutive equations to describe magnetoelasticity have been presented in Sec.~\ref{sec:theory} in a general form. The magnetization dynamics are described by Eq.~\eqref{eq:llg}, which includes the effective magnetic field in Eq.~\eqref{eq:h_eff}. In this paper, we consider an effective field that consists of one static and three dynamic contributions: a uniform static Zeeman field $\bm{H}_0$ along $\bm{\hat{x}}$ as well as dynamic dipolar, exchange, and magnetoelastic fields contributions, which can vary both in space and time. 

The dipolar field is found by solving Maxwell's equations. At microwave frequencies, the magnetostatic approximation can be applied since the wavelength of spin waves is typically much smaller than that of an electromagnetic wave at the same frequency. As a result, the generation of a magnetic field via the time-varying electric field can be neglected and the magnetic and electric fields become decoupled \cite{Gurevich96}. Within the magnetostatic approximation and in absence of electrical currents, the dipolar field is the solution of 
\begin{eqnarray}
\label{eq:Maxwell}
\nabla \cdot \bm{H}_\mathrm{d} &=& -\nabla \cdot \bm{M} \\
\nabla \times \bm{H}_\mathrm{d} &=& 0\,.
\end{eqnarray}
The solution can be written as \cite{Gurevich96}
\begin{equation}
\label{eq:Hdip_gen}
\mathbf{H}_{\mathrm{d}} = \frac{1}{4\pi} \int_{V'} \bar{D}(\mathbf{r}-\mathbf{r}')\mathbf{M}(\mathbf{r}') dV'\,,
\end{equation}
\noindent with $V'$ the volume of the magnetic material and $\bar{D}(\mathbf{r}-\mathbf{r}')$ the tensorial magnetostatic Green's function given by 
\begin{equation}
\bar{D}(\mathbf{r}-\mathbf{r}') = - \nabla_\mathbf{r} \nabla_\mathbf{r'} \frac{1}{|\mathbf{r}-\mathbf{r}'|}\,.
\end{equation} 
\noindent The general form of the exchange field can be deduced from Eqs.~ \eqref{eq:E_ex} and \eqref{eq:h_eff}. It can be written as 
\begin{equation}
\label{eq:Hex_gen}
\mathbf{H}_{\mathrm{ex}} = \frac{2A_{\mathrm{ex}}}{\mu_0 M_\mathrm{s}^2} \Delta \mathbf{M} = l_{\mathrm{ex}}^2 \Delta\mathbf{M} \equiv \lambda_{\mathrm{ex}} \Delta \mathbf{M}\,,
\end{equation}
\noindent with $\Delta$ the Laplace operator and $l_{\mathrm{ex}}$ the exchange length. For the magnetoelastic field, Eqs.~\eqref{eq:E_mel} and \eqref{eq:h_eff} result in 
\begin{equation}
\label{eq:Hmel_gen}
\mathbf{H}_{\mathrm{mel}} = -\frac{1}{\mu_0} \frac{\delta E_\mathrm{mel}}{\delta\mathbf{M}} = -\frac{2}{\mu_0M_\mathrm{s}}  \begin{bmatrix}
B_1\varepsilon_{\mathrm{xx}}m_x + B_2(\varepsilon_{\mathrm{xy}}m_y+\varepsilon_{zx}m_z) \\
B_1\varepsilon_{\mathrm{yy}}m_y + B_2(\varepsilon_{\mathrm{xy}}m_x+\varepsilon_{yz}m_z) \\
B_1\varepsilon_{\mathrm{zz}}m_z + B_2(\varepsilon_{\mathrm{zx}}m_x+\varepsilon_{yz}m_y) 
\end{bmatrix}\,,
\end{equation}
\noindent with $m_i$ the normalized magnetization components.

The elastodynamics are determined by the mechanical body forces, which are given by Eq.~\eqref{eq:bf}. Here, the total body force is the sum of the magnetoelastic and elastic forces. For materials with cubic (or higher, including isotropic) symmetry, these forces are given by
\begin{equation}
\label{eq:Fmel_gen}
\mathbf{F}_\mathrm{mel} = 2 B_1 \begin{bmatrix}
m_x \frac{\partial m_x}{\partial x} \\ m_y \frac{\partial m_y}{\partial y} \\ m_z \frac{\partial m_z}{\partial z} \end{bmatrix} + B_2 \begin{bmatrix}
m_x\left( \frac{\partial m_y}{\partial y} + \frac{\partial m_z}{\partial z} \right)  + m_y \frac{\partial m_x}{\partial y} + m_z \frac{\partial m_x}{\partial z} \\
m_y\left( \frac{\partial m_x}{\partial x} + \frac{\partial m_z}{\partial z} \right) + m_x \frac{\partial m_y}{\partial x} + m_z \frac{\partial m_y}{\partial z}  \\
m_z\left( \frac{\partial m_x}{\partial x} + \frac{\partial m_y}{\partial y} \right)  + m_x \frac{\partial m_z}{\partial x} + m_y \frac{\partial m_z}{\partial y} 
\end{bmatrix}
\end{equation}
\noindent and
\begin{equation}
\label{eq:Fel_gen}
\bm{F}_\mathrm{el} = \begin{bmatrix}
C_{11} \frac{\partial^2 u_x}{\partial x^2} + C_{44} \left( \frac{\partial^2 u_x}{\partial y^2} + \frac{\partial^2 u_x}{\partial z^2}  \right) + (C_{12} + C_{44}) \left( \frac{\partial^2 u_y}{\partial x \partial y} +\frac{\partial^2 u_z}{\partial x \partial z} \right)\\
C_{11} \frac{\partial^2 u_y}{\partial y^2} + C_{44} \left( \frac{\partial^2 u_y}{\partial x^2} + \frac{\partial^2 u_y}{\partial z^2}  \right) + (C_{12} + C_{44}) \left(  \frac{\partial^2 u_x}{\partial x \partial y}+ \frac{\partial^2 u_z}{\partial z \partial y} \right) \\
C_{11} \frac{\partial^2 u_z}{\partial z^2} + C_{44} \left( \frac{\partial^2 u_z}{\partial x^2} +  \frac{\partial^2 u_z}{\partial y^2} \right) + (C_{12} + C_{44}) \left(  \frac{\partial^2 u_x}{\partial x \partial z}+ \frac{\partial^2 u_y}{\partial z \partial y} \right)
\end{bmatrix}\,,
\end{equation}
\noindent where $C_\mathrm{ij}$ are the components of the stiffness tensor.

For a waveguide with the static magnetization along the wave propagation direction, these fields and forces can be further simplified. The geometry is represented in Fig.~\ref{fig:geometry}, with the propagation direction along $\bm{\hat{x}}$ and the out-of-plane direction along the $\bm{\hat{z}}$. The ansatz for a propagating magnetoelastic wave is described in Eq.~\eqref{eq:ansatz} for a known magnetization profile and an unknown displacement profile. In this geometry and with this ansatz, the dynamic dipolar field can then approximated by \cite{Damon60,Harte68}
\begin{equation}
\label{hdip}
	\bm{h}_\mathrm{d} = -\begin{bmatrix}
	0 \\ P \frac{\kappa_n^2}{k_\mathrm{tot}^2} \\ 1-P
	\end{bmatrix}\bm{M}\,,
\end{equation}
\noindent with 
\begin{equation}
\label{eq:P}
P = 1-\frac{1-e^{-k_\mathrm{tot}d}}{k_\mathrm{tot}d} \,,
\end{equation}
\noindent $k_\mathrm{tot}^2 = k_x^2+\kappa_n^2$, $d$ the waveguide thickness, and $\kappa_n=\frac{n\pi}{w_\mathrm{eff}}$, with $w_\mathrm{eff}$ the effective waveguide width. The dynamic exchange field is given by
\begin{equation}
\label{hex}
\mathbf{h}_{\mathrm{ex}} = -\lambda_{\mathrm{ex}} k_\mathrm{tot}^2 \bm{M}
\end{equation}
\noindent and the dynamic magnetoelastic field becomes 
\begin{equation}
\label{eq:hmel_guide}
\mathbf{h}_{\mathrm{mel}} =  -\frac{1}{\mu_0M_\mathrm{s}}  \begin{bmatrix}
2B_1ik_xu_xm_x + B_2\left( \left(\frac{\partial u_x}{\partial y}+ik_xu_y \right) m_y+ik_x u_z m_z\right) \\
2B_1\frac{\partial u_y}{\partial y}m_y + B_2\left( \left(\frac{\partial u_x}{\partial y}+ik_xu_y \right)m_x+\frac{\partial u_z}{\partial y} m_z\right) \\
  B_2\left( ik_xu_zm_x +\frac{\partial u_z}{\partial y} m_y\right)
\end{bmatrix}\,.
\end{equation}
\noindent Analogously, the magnetoelastic and elastic body forces respectively can be written as
\begin{equation}
\label{eq:Fmel_guide}
\mathbf{f}_\mathrm{mel} = 2 B_1 \begin{bmatrix}
m_x \frac{\partial m_x}{\partial x} \\ m_y \frac{\partial m_y}{\partial y} \\0 \end{bmatrix} + B_2 \begin{bmatrix}
 m_x\frac{\partial m_y}{\partial y} + m_y\frac{\partial m_x}{\partial y}   \\ 
 m_y \frac{\partial m_x}{\partial x} + ik_x m_x m_y   \\
 m_z \left(\frac{\partial m_x}{\partial x} + \frac{\partial m_y}{\partial y}  \right) +ik_xm_xm_z + m_y \frac{\partial m_z}{\partial y} \end{bmatrix} 
\end{equation}
\noindent and
\begin{equation}
\label{eq:Fel_guide}
\bm{f}_\mathrm{el} = \begin{bmatrix}
-C_{11} k_x^2 u_x + C_{44} \frac{\partial^2 u_x}{\partial y^2} + (C_{12} + C_{44})ik_x \frac{\partial u_y}{ \partial y} \\
C_{11} \frac{\partial^2 u_y}{\partial y^2} + - C_{44} k_x^2 u_y + (C_{12} + C_{44})ik_x \frac{\partial u_x}{ \partial y} \\
C_{44} \left( -ik_xu_z +  \frac{\partial^2 u_z}{\partial y^2} \right) 
\end{bmatrix}\,.
\end{equation}

Substituting these terms in the equations of motion \eqref{eq:llg} and \eqref{eq:elastodynamic} and neglecting damping as well as second order terms results in linearized equations of motion for the displacement and magnetization, which can be written as
\begin{equation}
\label{eq:system_of_melwaves}
\begin{aligned}
-\rho \omega^2 u_x &= -C_{11} k_x^2 u_x + C_{44} \partial_y^2 u_x  + (C_{12} + C_{44})ik_x \partial_y  u_y + \frac{B_2}{M_\mathrm{s}}  i\kappa_n m_y \\
-\rho \omega^2 u_y &= C_{11} \partial_y^2 u_y - C_{44}   k_x^2 u_y + (C_{12} + C_{44}) ik_x \partial_y  u_x + \frac{B_2}{M_\mathrm{s}} ik_xm_y \\
-\rho \omega^2 u_z &= -C_{44} \left( k_x^2 u_z -\partial_y^2 u_z\right) + \frac{B_2}{M_\mathrm{s}} ik_xm_z\\
i\omega m_y &= -\omega_\mathrm{mz}m_z - \gamma B_2 ik_x u_z  \\
i\omega m_z &= \omega_\mathrm{my}m_y + \gamma B_2 \left(i k_x u_y + \partial_y u_x \right)\,.
\end{aligned}
\end{equation}
Reordering the terms leads to
\begin{equation}
\begin{bmatrix}
v_\parallel^2k_x^2 - v_\perp^2 \partial_y^2 - \omega^2  &  -iv_\asymp^2k_x\partial_y  & 0 & \frac{i\kappa_nB_2}{\rho M_\mathrm{s}} & 0 \\
-iv_\asymp^2k_x\partial_y  & v_\perp^2 k_x^2 -v_\parallel^2 \partial_y^2 - \omega^2 & 0 & \frac{ik_xB_2}{\rho M_\mathrm{s}}  & 0 \\
0 & 0 & v_\perp^2 \left( k_x^2 -\partial_y^2 \right)-\omega^2  & 0 & \frac{ik_xB_2}{\rho M_\mathrm{s}}  \\
\gamma B_2 \partial_y & \gamma i B_2 k_x  & 0 & \omega_\mathrm{my} &  -i\omega \\
0 & 0 & \gamma B_2 ik_x & i\omega& \omega_\mathrm{mz}
\end{bmatrix}\bm{w}_n(x,y,t) = \begin{bmatrix} 0\\0\\0\\0\\0 \end{bmatrix}\, ,
\end{equation}

\noindent with $v_\parallel^2=C_{11}/\rho$, $v_\perp^2=C_{44}/\rho$, $v_\asymp^2 = (C_{12}+C_{44})/\rho$, $C_{ij}$ the stiffness constants, $\partial_y=\partial/\partial y$,
\begin{eqnarray}
\omega_\mathrm{my} &=& \omega_0 + \omega_\mathrm{M} \left(\lambda_{\mathrm{ex}}k_\mathrm{tot}^2 + P\frac{\kappa_n^2}{k_\mathrm{tot}^2} \right) \,,\\
\omega_\mathrm{mz} &=& \omega_0+\omega_\mathrm{M} (\lambda_{\mathrm{ex}} k_\mathrm{tot}^2 +1-P)\,.
\end{eqnarray}
	
\begin{acknowledgments}
This work has been supported by imec’s industrial affiliate program on beyond-CMOS
logic. It has also received funding from the European Union’s Horizon 2020 research and innovation program within the FET-OPEN project CHIRON under grant agreement No.
801055. F.V. acknowledges financial support from the Research Foundation -- Flanders
(FWO) through grant No.~1S05719N. J.L. is supported by the Research Foundation -- Flanders (FWO) through a postdoctoral fellowship.
\end{acknowledgments}

\bibliography{refs5.bib}

\end{document}